\documentclass[aps,prd,twocolumn,superscriptaddress,showpacs,preprintnumbers,amsmath,amssymb,floatfix]{revtex4-2}

\usepackage{graphicx}
\usepackage{dcolumn}
\usepackage{bm}
\usepackage{hyperref}
\usepackage{caption}
\usepackage{float}
\usepackage{subcaption}
\usepackage{tabularx}

\usepackage[mathlines]{lineno}

\begin{document}


\title{\quad\\[1.1cm]Measurement of the Resonant and Non-Resonant Branching Ratios in \texorpdfstring{$\Xi_{c}^{0} \rightarrow \Xi^{0} K^+ K^-$}{Xic0XiKK} }

\noaffiliation
\affiliation{University of the Basque Country UPV/EHU, 48080 Bilbao}
\affiliation{University of Bonn, 53115 Bonn}
\affiliation{Brookhaven National Laboratory, Upton, New York 11973}
\affiliation{Budker Institute of Nuclear Physics SB RAS, Novosibirsk 630090}
\affiliation{Faculty of Mathematics and Physics, Charles University, 121 16 Prague}
\affiliation{Chonnam National University, Gwangju 61186}
\affiliation{University of Cincinnati, Cincinnati, Ohio 45221}
\affiliation{Deutsches Elektronen--Synchrotron, 22607 Hamburg}
\affiliation{University of Florida, Gainesville, Florida 32611}
\affiliation{Key Laboratory of Nuclear Physics and Ion-beam Application (MOE) and Institute of Modern Physics, Fudan University, Shanghai 200443}
\affiliation{Gifu University, Gifu 501-1193}
\affiliation{SOKENDAI (The Graduate University for Advanced Studies), Hayama 240-0193}
\affiliation{Gyeongsang National University, Jinju 52828}
\affiliation{Department of Physics and Institute of Natural Sciences, Hanyang University, Seoul 04763}
\affiliation{University of Hawaii, Honolulu, Hawaii 96822}
\affiliation{High Energy Accelerator Research Organization (KEK), Tsukuba 305-0801}
\affiliation{J-PARC Branch, KEK Theory Center, High Energy Accelerator Research Organization (KEK), Tsukuba 305-0801}
\affiliation{Higher School of Economics (HSE), Moscow 101000}
\affiliation{Forschungszentrum J\"{u}lich, 52425 J\"{u}lich}
\affiliation{IKERBASQUE, Basque Foundation for Science, 48013 Bilbao}
\affiliation{Indian Institute of Technology Bhubaneswar, Satya Nagar 751007}
\affiliation{Indian Institute of Technology Hyderabad, Telangana 502285}
\affiliation{Indian Institute of Technology Madras, Chennai 600036}
\affiliation{Indiana University, Bloomington, Indiana 47408}
\affiliation{Institute of High Energy Physics, Chinese Academy of Sciences, Beijing 100049}
\affiliation{Institute of High Energy Physics, Vienna 1050}
\affiliation{Institute for High Energy Physics, Protvino 142281}
\affiliation{INFN - Sezione di Napoli, 80126 Napoli}
\affiliation{INFN - Sezione di Torino, 10125 Torino}
\affiliation{Advanced Science Research Center, Japan Atomic Energy Agency, Naka 319-1195}
\affiliation{J. Stefan Institute, 1000 Ljubljana}
\affiliation{Institut f\"ur Experimentelle Teilchenphysik, Karlsruher Institut f\"ur Technologie, 76131 Karlsruhe}
\affiliation{Kennesaw State University, Kennesaw, Georgia 30144}
\affiliation{Department of Physics, Faculty of Science, King Abdulaziz University, Jeddah 21589}
\affiliation{Kitasato University, Sagamihara 252-0373}
\affiliation{Korea Institute of Science and Technology Information, Daejeon 34141}
\affiliation{Korea University, Seoul 02841}
\affiliation{Kyoto Sangyo University, Kyoto 603-8555}
\affiliation{Kyungpook National University, Daegu 41566}
\affiliation{Universit\'{e} Paris-Saclay, CNRS/IN2P3, IJCLab, 91405 Orsay}
\affiliation{P.N. Lebedev Physical Institute of the Russian Academy of Sciences, Moscow 119991}
\affiliation{Faculty of Mathematics and Physics, University of Ljubljana, 1000 Ljubljana}
\affiliation{Ludwig Maximilians University, 80539 Munich}
\affiliation{Luther College, Decorah, Iowa 52101}
\affiliation{Malaviya National Institute of Technology Jaipur, Jaipur 302017}
\affiliation{University of Maribor, 2000 Maribor}
\affiliation{Max-Planck-Institut f\"ur Physik, 80805 M\"unchen}
\affiliation{School of Physics, University of Melbourne, Victoria 3010}
\affiliation{University of Mississippi, University, Mississippi 38677}
\affiliation{University of Miyazaki, Miyazaki 889-2192}
\affiliation{Graduate School of Science, Nagoya University, Nagoya 464-8602}
\affiliation{Universit\`{a} di Napoli Federico II, 80126 Napoli}
\affiliation{Nara Women's University, Nara 630-8506}
\affiliation{National Central University, Chung-li 32054}
\affiliation{National United University, Miao Li 36003}
\affiliation{Department of Physics, National Taiwan University, Taipei 10617}
\affiliation{H. Niewodniczanski Institute of Nuclear Physics, Krakow 31-342}
\affiliation{Nippon Dental University, Niigata 951-8580}
\affiliation{Niigata University, Niigata 950-2181}
\affiliation{Novosibirsk State University, Novosibirsk 630090}
\affiliation{Osaka City University, Osaka 558-8585}
\affiliation{Pacific Northwest National Laboratory, Richland, Washington 99352}
\affiliation{Peking University, Beijing 100871}
\affiliation{University of Pittsburgh, Pittsburgh, Pennsylvania 15260}
\affiliation{Research Center for Nuclear Physics, Osaka University, Osaka 567-0047}
\affiliation{Meson Science Laboratory, Cluster for Pioneering Research, RIKEN, Saitama 351-0198}
\affiliation{Department of Modern Physics and State Key Laboratory of Particle Detection and Electronics, University of Science and Technology of China, Hefei 230026}
\affiliation{Seoul National University, Seoul 08826}
\affiliation{Showa Pharmaceutical University, Tokyo 194-8543}
\affiliation{Soochow University, Suzhou 215006}
\affiliation{Soongsil University, Seoul 06978}
\affiliation{Sungkyunkwan University, Suwon 16419}
\affiliation{School of Physics, University of Sydney, New South Wales 2006}
\affiliation{Department of Physics, Faculty of Science, University of Tabuk, Tabuk 71451}
\affiliation{Tata Institute of Fundamental Research, Mumbai 400005}
\affiliation{Department of Physics, Technische Universit\"at M\"unchen, 85748 Garching}
\affiliation{School of Physics and Astronomy, Tel Aviv University, Tel Aviv 69978}
\affiliation{Toho University, Funabashi 274-8510}
\affiliation{Earthquake Research Institute, University of Tokyo, Tokyo 113-0032}
\affiliation{Department of Physics, University of Tokyo, Tokyo 113-0033}
\affiliation{Tokyo Institute of Technology, Tokyo 152-8550}
\affiliation{Tokyo Metropolitan University, Tokyo 192-0397}
\affiliation{Utkal University, Bhubaneswar 751004}
\affiliation{Virginia Polytechnic Institute and State University, Blacksburg, Virginia 24061}
\affiliation{Wayne State University, Detroit, Michigan 48202}
\affiliation{Yamagata University, Yamagata 990-8560}
\affiliation{Yonsei University, Seoul 03722}
  \author{J.~T.~McNeil}\affiliation{University of Florida, Gainesville, Florida 32611} 
  \author{J.~Yelton}\affiliation{University of Florida, Gainesville, Florida 32611} 
  \author{J.~Bennett}\affiliation{University of Mississippi, University, Mississippi 38677} 
  \author{I.~Adachi}\affiliation{High Energy Accelerator Research Organization (KEK), Tsukuba 305-0801}\affiliation{SOKENDAI (The Graduate University for Advanced Studies), Hayama 240-0193} 
  \author{K.~Adamczyk}\affiliation{H. Niewodniczanski Institute of Nuclear Physics, Krakow 31-342} 
  \author{J.~K.~Ahn}\affiliation{Korea University, Seoul 02841} 
  \author{H.~Aihara}\affiliation{Department of Physics, University of Tokyo, Tokyo 113-0033} 
  \author{S.~Al~Said}\affiliation{Department of Physics, Faculty of Science, University of Tabuk, Tabuk 71451}\affiliation{Department of Physics, Faculty of Science, King Abdulaziz University, Jeddah 21589} 
  \author{D.~M.~Asner}\affiliation{Brookhaven National Laboratory, Upton, New York 11973} 
  \author{H.~Atmacan}\affiliation{University of Cincinnati, Cincinnati, Ohio 45221} 
% ...
\collaboration{The Belle Collaboration}

\begin{abstract}
Using the entire data sample of $980$ $\rm fb^{-1}$ integrated luminosity collected with the Belle detector at the KEKB asymmetric-energy $e^{+}e^{-}$ collider, we present an amplitude analysis measuring the branching fractions of the Cabibbo-allowed, $W$-exchange resonant decay $\Xi_{c}^{0} \rightarrow \Xi^{0} \phi (\to K^+ K^-)$ with a polarized $\phi$ and the non-resonant decay via a direct process $\Xi_{c}^{0} \rightarrow \Xi^{0} K^+ K^-$. We present these measurements, relative to the normalization mode $\Xi^{-}\pi^{+}$, and find branching ratios $\frac{\mathcal{B}(\Xi_{c}^{0} \rightarrow \Xi^{0} \phi (\rightarrow K^{+}K^{-}))}{\mathcal{B}(\Xi_{c}^{0} \rightarrow \Xi^{-} \pi^{+})} \! = \! 0.036 \pm 0.004\,(\rm stat.)\pm0.002\,(\rm syst.)$ and $\frac{\mathcal{B}(\Xi_{c}^{0} \rightarrow \Xi^{0} K^{+} K^{-})}{\mathcal{B}(\Xi_{c}^{0} \rightarrow \Xi^{-} \pi^{+})} \! = \! 0.039 \pm 0.004\,(\rm stat.)\pm0.002\,(\rm syst.)$ which suggest that only minor cusping peaks occur in the combinatorial background of $\Omega^{*-} \to \Xi^{0}K^{-}$ due to these $\Xi_{c}^{0}$ decays.
\end{abstract}

\maketitle


\section{\label{sec:S1}Introduction}

\begin{figure}[b]
    \label{fig:F1}
    
    \textbf{(a)}
         \includegraphics[height=0.165\textheight,width=0.86\linewidth]{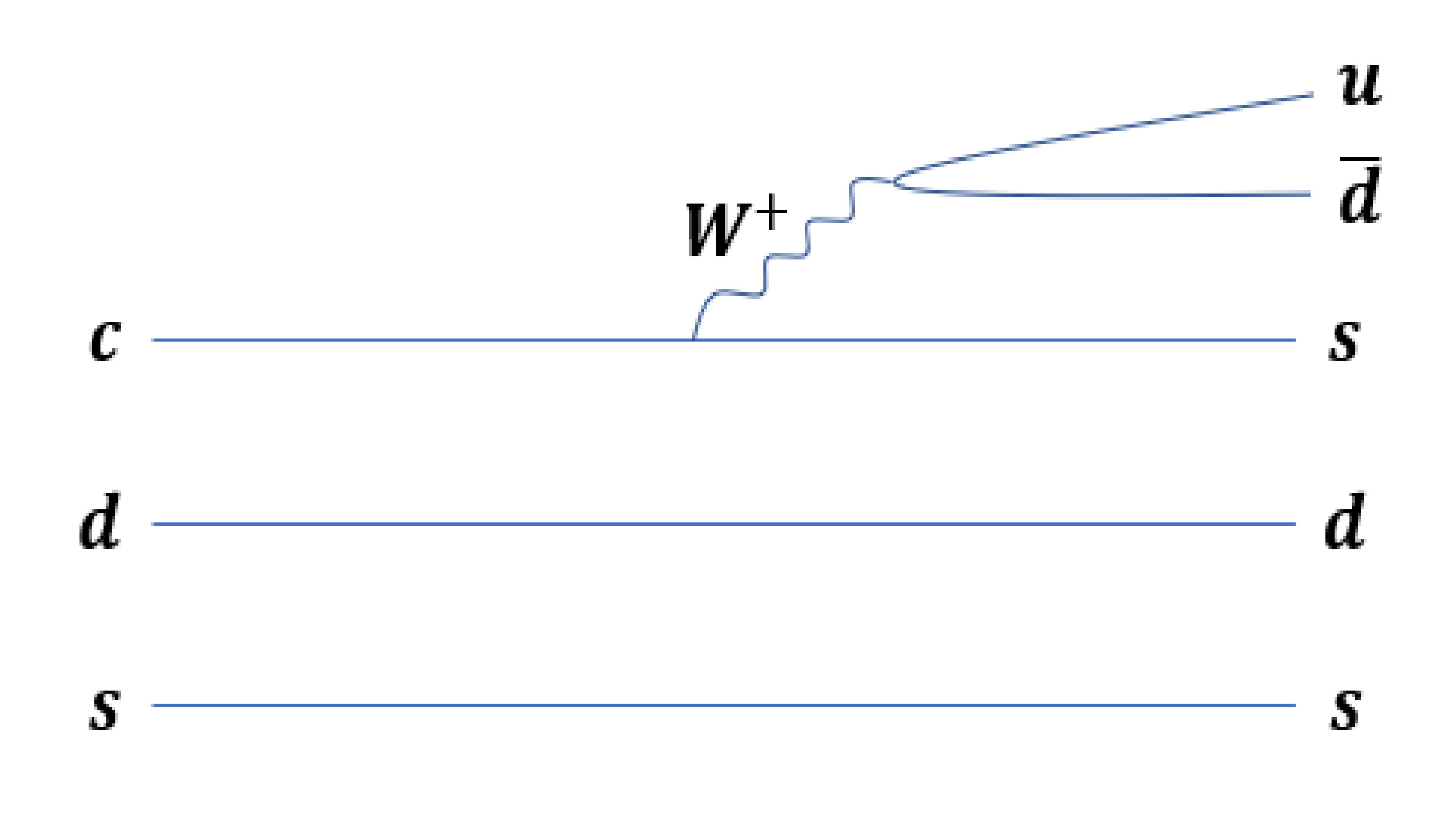}
    
    \vspace{0.5cm}
    
    \textbf{(b)}
        \includegraphics[height=0.155\textheight,width=0.81\linewidth]{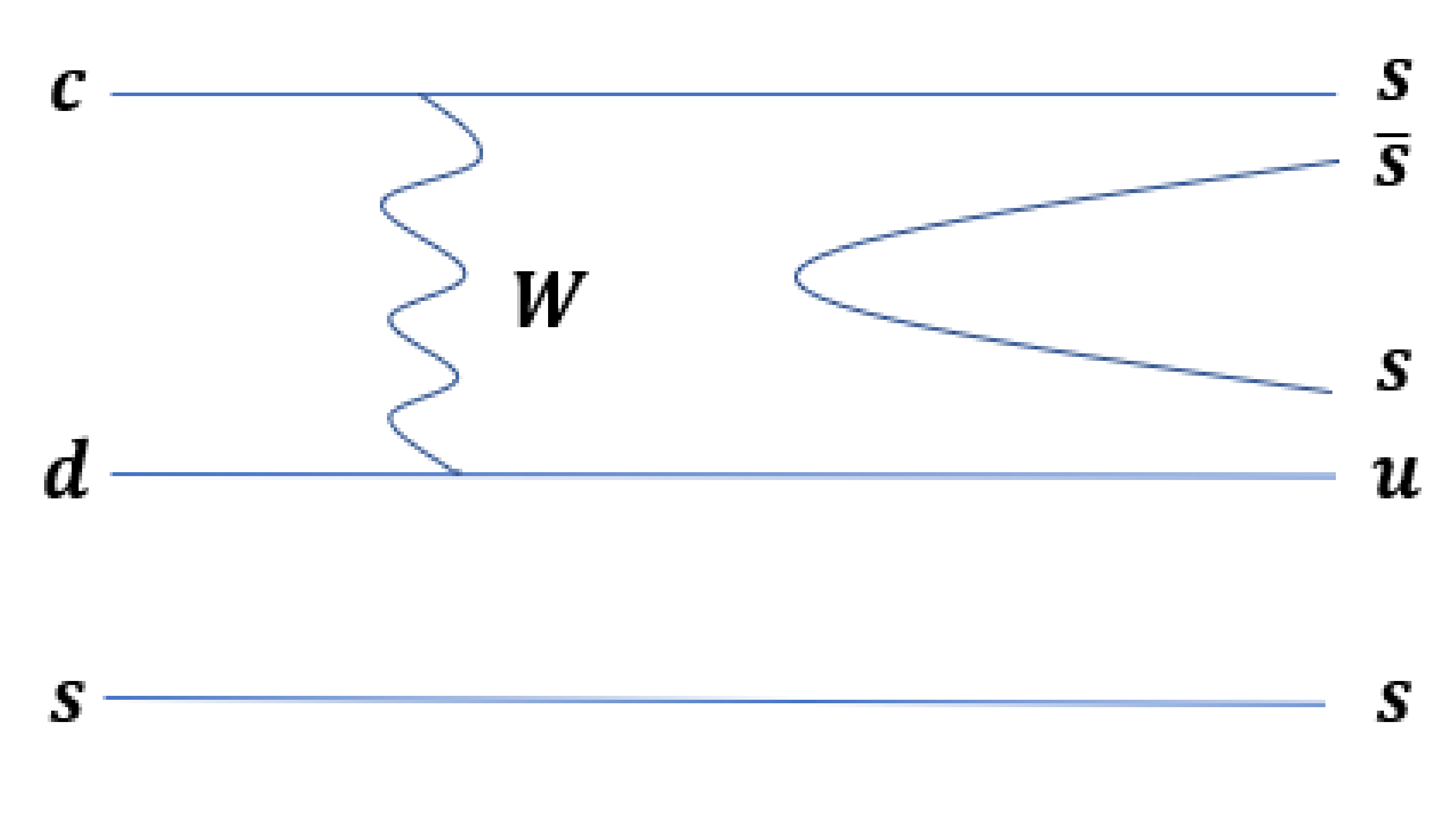}
    \caption{Decay diagrams depicting the non-leptonic, weak decay of $\Xi_{c}^{0} \rightarrow \Xi^{-} \pi^{+}$ via an emitted $W$-boson \textbf{(a)} and the Cabibbo-allowed $W$-exchange, $s\bar{s}$-popping decay of the $\Xi_{c}^{0} \to \Xi^{0} K^{+} K^{-}$ which can resonate through $\phi \to K^{+} K^{-}$ \textbf{(b)}.}
\end{figure}

The most simple model of the $\Xi_{c}^{0}(dsc)$ baryon non-leptonic decay occurs via $c\!\to\!s$ transitions into $\Xi^{-}\pi^{+}$ by way of an emitted $W$-boson. However, these decays may also occur via charged current ($W$-exchange) interactions between the quarks of the $\Xi_{c}^{0}$ baryon. The most probable example of this decay is the Cabibbo-allowed $W$-exchange between the $cd$ quarks, as shown in the decay diagrams of \hyperref[fig:F1]{Fig. 1}. In this type of decay the $cd$ quarks exchange charge via the interacting $W$-boson and transition into $su$ quarks respectively. Generally the remaining momentum from this decay escapes via an emitted gluon which immediately produces a quark-antiquark pair. For this work, we study the case in which the emitted gluon from the $W$-exchange decays into an $s\bar{s}$ quark pair ($s\bar{s}$-popping) through $cd\!\to\!W^{+}\!\to\!su(g\!\to\!s\bar{s})$.

Such $s\bar{s}$-popping decays are in general suppressed compared with the analogous light-quark popping decays but have more cleanly measurable final states. Other $s\bar{s}$-popping decays have been measured, and in particular by the Belle Collaboration the similar but Cabibbo-suppressed mode $\Xi_c^0\!\to\!\Lambda K^+ K^-$ \cite{suppressed}. In this work we study the previously unobserved \cite{PDG} Cabibbo-allowed mode $\Xi_{c}^{0}\!\to\!\Xi^{0} K^{+} K^{-}$ and the fraction of this decay that resonates through $\phi\!\to\!K^{+} K^{-}$. 

One particular motivation for the study of these new $\Xi_{c}^{0}$ decay channels is due to the recently discovered excited $\Omega^{-}$ baryon at Belle \cite{excitedOm}. This excited $\Omega^{-}$ was found in the $\Xi^{0}K^{-}$ channel with an invariant mass of $2.012$ GeV where for this work we use natural units with $c = 1$. From quark model predictions of heavy baryon excited states, there are good theoretical reasons to expect that this $\Omega^{-}$ baryon may have a partner near $1.95$ GeV \cite{excitedOm,Shen} and low-statistics indications of an excess in this region can be inferred. However, it is suspected that a $\Xi_{c}^{0}$ decaying to $\Xi^{0}K^+K^-$ through a polarized $\phi\!\to\!K^{+}K^{-}$ could produce peaks in the $\Xi^{0}K^{\pm}$ invariant mass spectra as well. These topological substructures are due to the helicity angles of the $\Xi_{c}^{0}$ polarizing the $\phi$ in the $1/2\!\to\!1/2 + 1$ resonant decay process \cite{Lambdab}. Hence, the decay substructure of the $\Xi_{c}^{0} \to \Xi^{0} K^{+} K^{-}$ must be studied to assure that any evidence of an excess in this region can be attributed to an excited $\Omega^{-}$ candidate and not an artifact of these resonant decays.

The Belle detector and KEKB asymmetric $e^{+}e^{-}$ collider collected the world's largest sample of $\Upsilon(4S)$ energy data over a $1999\!-\!2010$ run period with a total $980$ $\rm fb^{-1}$ integrated luminosity for analysis across all energy ranges \cite{PTEP}. The Belle detector was a large, asymmetric detector hermetically consisting of six subdetectors. From inner-to-outermost subdetector, the Belle detector included the following components.

For charged particle tracking, four innermost layers of double-sided silicon strip vertex detectors surrounded the beryllium beam pipe encased by a 50-layer central drift chamber (CDC). These tracking detectors were asymmetrically oriented in the $z$-axis to assure optimal solid angle coverage with respect to the interaction point (IP). For charged particle identification (PID), defined $\mathcal{L}(x\!:\!y)\!=\!\mathcal{L}_{x} / (\mathcal{L}_{x}\!+\!\mathcal{L}_{y})$ for likelihoods $\mathcal{L}_i$ of identifying the charged particles $p$, $\pi$, and $K$; along with $dE/dx$ measurements in the CDC, Belle included two subdetectors. The aerogel Cherenkov counters (ACC) were located along the barrel and outer, large solid angle of the CDC, and the time-of-flight counters were positioned just outside of the ACC with respect to the barrel. For electron and $\gamma$ detection via electromagnetic showers, along the entire Belle detector $23^{o}\!<\!\theta\!<\!139^{o}$ solid angle, CsI(Tl) crystals comprised the electromagnetic calorimeter (ECL). Outside the $1.5$ T solenoid coil, the remaining subdetectors included layered resistive plate counters with iron yoke for muon and $K_L$ detection \cite{belle,Physics}. 

\section{\label{sec:S2}Reconstruction}
Event reconstruction for this analysis is performed entirely in the Belle II Software Framework by converting Belle data structures to Belle II data structures \cite{basf2,b2bii}. Background to the reconstructed invariant mass distribution of the  $\Xi^{0} K^{+} K^{-}$ signal is in large part due to combinatorics and clone hyperon reconstructions from soft and overlapping $\gamma$ candidates. Hence, it is necessary to reconstruct $\Xi^{0} \to \Lambda \pi^0$ candidates with a good signal-to-background ratio. We detail the reconstruction procedure and selection criteria for the $\Xi_c^0$ modes: $\Xi_c^0\!\to\!(\Xi^-\!\to\!(\Lambda\!\to\!p\pi^-)\pi^-)\pi^+$ and $\Xi_c^0\!\to\!(\Xi^0\!\to\!(\Lambda\!\to\!p\pi^-)(\pi^0\!\to\!\gamma\gamma))K^+K^-$, by following a similar methodology to that of the $\Omega(2012) \to \Xi^0 K^-$ observation \cite{excitedOm} which used the previous Belle Software Framework. 

Preliminary $\Lambda$ candidates must exhibit the expected kinematics from a long-lived hyperon decay, $\Xi \to \Lambda \pi$. These $\Lambda \to p \pi^-$ candidates are found with a local vertex reconstruction using a Kalman-Filter \cite{treefitter} and selected with the following kinematics: a reconstructed mass in $\pm3.5$ MeV range of the nominal mass \cite{PDG} which is approximately $99\%$ efficient; a $\cos(\alpha_{xyz}) > 0.0$ in the 3D plane; a distance of the decay vertex with respect to the IP greater than $0.35$ cm; and a loose PID requirement on the $p$ track with $\mathcal{L}(p\!:\!\pi)$ and $\mathcal{L}(p\!:\!K)$ greater than $0.2$ which is approximately $99\%$ efficient. For these kinematics we define $\alpha_{xyz}$ $(\alpha_{xy})$ as the 3D (2D) angle between a vector from the IP to the decay vertex and the momentum vector at the decay vertex. All hyperons with higher strangeness are then reconstructed with the described $\Lambda$ candidates and all good $\pi^-$ or $\pi^0 \to \gamma \gamma$ candidates a priori selected by internal Belle studies of the CDC and ECL performance \cite{b2bii}. The decay vertex of these hyperon candidates is then globally reconstructed using a decay chain fitter with mass-constrained daughters to improve the signal-to-background \cite{basf2,treefitter}. 

\begin{figure}[ht!]
    \label{fig:F2}
    \includegraphics[height=0.24\textheight,width=0.75\linewidth]{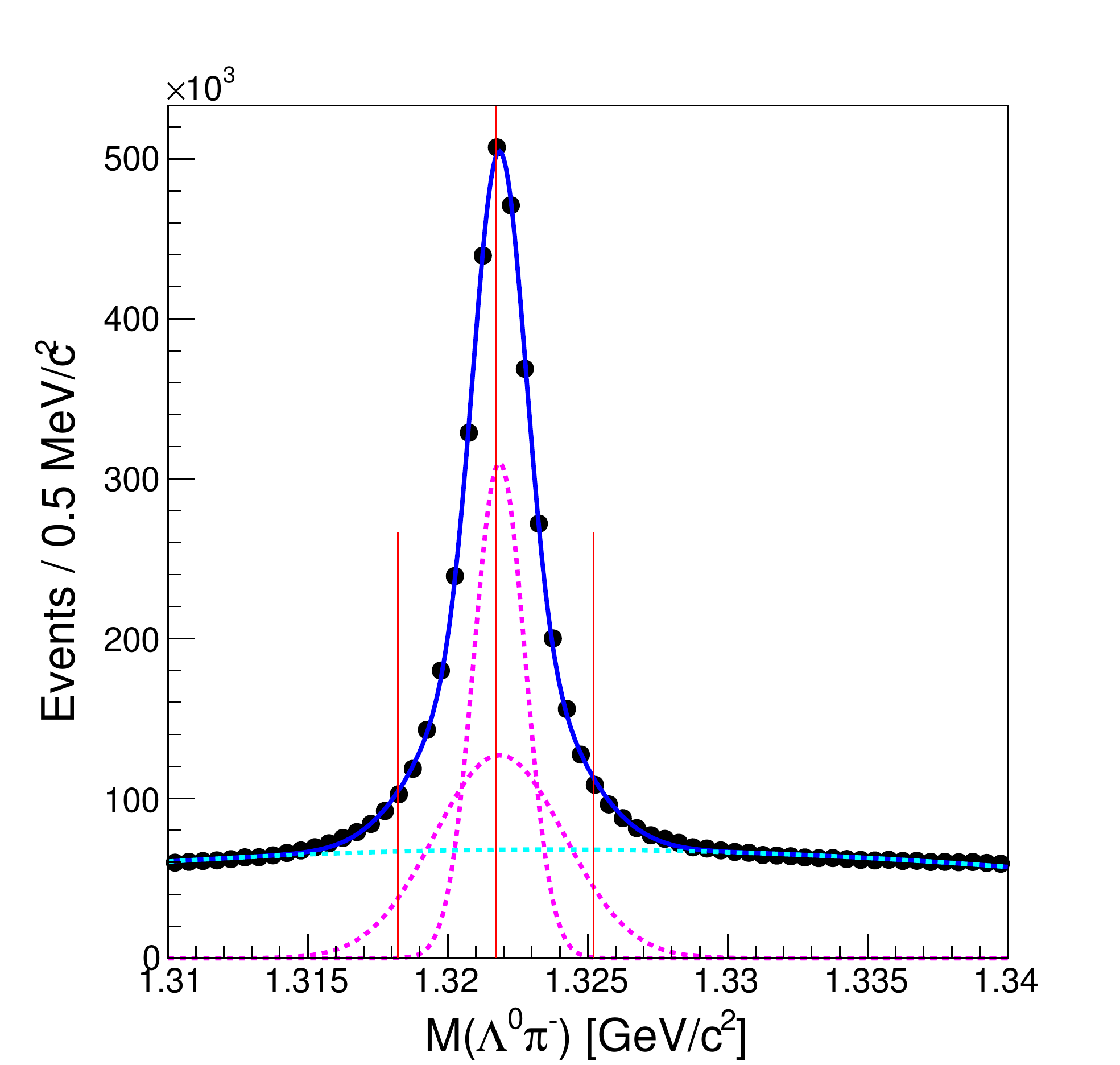}
    \par\medskip
    \includegraphics[height=0.24\textheight,width=0.75\linewidth]{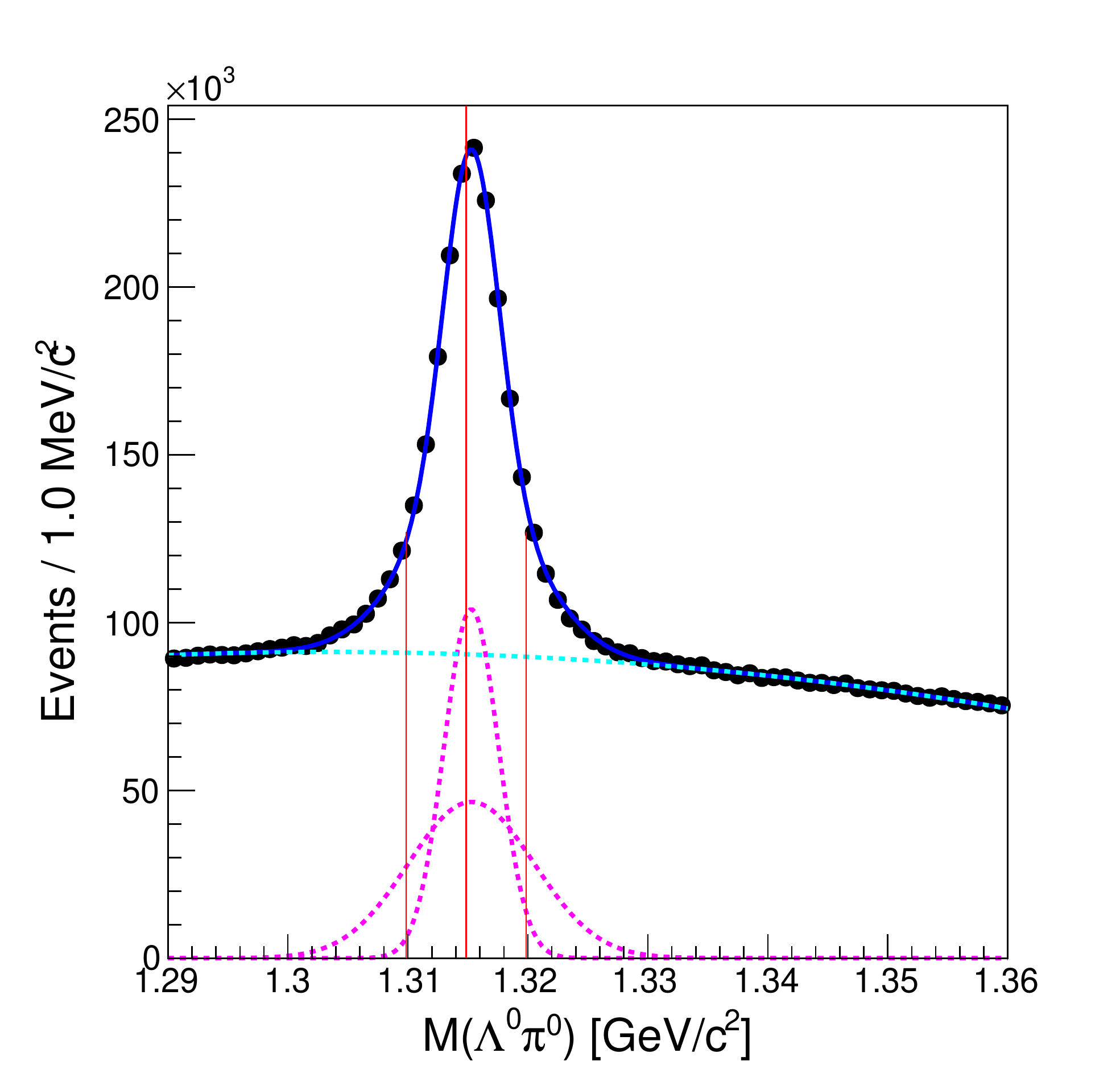}
    \caption{Selected signal bands (marked in red half-lines) with respect to the nominal masses (marked in red full-lines) of the invariant mass distributions for the hyperon decays $\Xi^- \to \Lambda \pi^-$ (upper) and $\Xi^0 \to \Lambda \pi^0$ (lower) in the data sample (black) with an unbinned likelihood fit to a double guassian PDF (solid blue) with Chebyshev polynomial background (dashed cyan) and each individual Gaussian contribution to the signal PDF (dashed magenta).}
\end{figure}

Optimal $\Xi^- \to \Lambda \pi^-$ selection is less crucial for the normalization mode decay due to the high yield of these charmed baryon decays produced at Belle. We loosely select $\Xi^- \to \Lambda \pi^{-}$ candidates for this analysis with the following kinematics: a reconstructed invariant mass in $\pm3.5$ MeV range of the nominal mass \cite{PDG} which is approximately $4\sigma$ with respect to the primary resolution; a reconstructed decay vertex chi-squared probability consistent with a $\chi^2$ per degree of freedom less than $20$; a distance of the $\Xi^-$ decay vertex with respect to the IP preceding the $\Lambda$ decay vertex distance; a distance of the $\Xi^-$ decay vertex with respect to the IP greater than $0.1$ cm; a $\pi^-$ transverse momentum greater than $50$ MeV; and a loose ratio between the $\Lambda$ and $\Xi^{-}$ hyperon $\cos(\alpha_{xy})$ angles in the tangential plane which is consistent with a hyperon decay. The reconstructed invariant mass for these $\Xi^- \to \Lambda \pi^-$ hyperons is shown in \hyperref[fig:F2]{Fig. 2} together with an unbinned maximum likelihood fit to the data using a probability density function (PDF) comprised of a double Gaussian
signal function and a second-order Chebyshev polynomial background. 

We reconstruct $\Xi^0 \to \Lambda \pi^0$ by refitting the $\pi^0$ due to the a priori candidates containing no directional information. Using a global decay chain reconstruction \cite{treefitter,basf2} the $\pi^0$ candidates are refit to the decay vertex of the $\Xi^0 \to \Lambda \pi^0$ with the $\Xi^0$ constrained to come from the nominal IP region. After refitting these $\pi^0$ candidates, the $\Xi^0$ hyperons used for this analysis are selected with the following kinematics: a reconstructed invariant mass in a $\pm5$ MeV range of the nominal mass \cite{PDG} which is approximately $2\sigma$ with respect to the primary resolution; a reconstructed decay vertex chi-squared probability consistent with a $\chi^2$ per degree of freedom less than $20$; a distance of the $\Xi^0$ decay vertex with respect to the IP preceding the $\Lambda$ decay vertex distance; a distance of the $\Xi^0$ decay vertex with respect to the IP greater than $1.4$ cm; a re-fit $\pi^0$ momentum greater than $150$ MeV; a re-fit $\pi^{0}$ mass range $\pm10.4$ MeV about the $\pi^{0}$ nominal mass \cite{PDG} which is approximately $2\sigma$; and lastly, a $\Lambda$ tangential $\alpha_{xy}$ deflection angle greater than the IP produced $\Xi^0$ angle of $\alpha_{xy} \approx 0$. The reconstructed invariant mass for these $\Xi^0 \to \Lambda \pi^0$ hyperons is shown in \hyperref[fig:F2]{Fig. \!2} together with an unbinned maximum likelihood fit to the data using a PDF comprised of a double Gaussian signal function and a second-order Chebyshev polynomial background.

For the normalization channel $\Xi_c^0\to\Xi^-\pi^+$, all $\Xi^-$ hyperons described above are mass-constrained and combined with selected $\pi^+$ candidates. These $\pi^+$ candidates are selected with a point of closest approach (POCA) in the $xy$-plane less than $0.2$ cm, a POCA along the $z$-axis less than $1.0$ cm, and a loose PID requirement with $\mathcal{L}(\pi\!:\!p)$ and $\mathcal{L}(\pi\!:\!K)$ greater than $0.2$ which is approximately $99\%$ efficient. To reconstruct the signal channel $\Xi_c^0\!\to\!\Xi^{0} K^{+} K^{-}$, the $\Xi^0$ hyperons described above are mass-constrained and combined with similarly selected $K^\pm$ candidates. These $K^\pm$ candidates are selected with a POCA in the $xy$-plane less than $0.2$ cm, a POCA along the $z$-axis less than $1.0$ cm, and a tight PID requirements with $\mathcal{L}(K\!:\!p)$ and $\mathcal{L}(K\!:\!\pi)$ greater than $0.9$ which is approximately $83\%$ efficient. The decay vertex of these $\Xi_c^0$ candidates in each mode is then reconstructed locally and constrained to the nominal IP profile region of the Belle detector. The final $\Xi_c^0$ candidates are then optimally selected via a figure of merit with a scaled momentum $x_p > 0.5$ for $x_p = p^*/\sqrt{E_{\rm beam}^2 - M_{\Xi_{c}^{0}}^{2}}$ where $p^*$ is the momentum in the $e^+e^-$ center of mass (CM) frame. This requirement is typically used to produce a good signal-to-background ratio while retaining high efficiency for charmed baryons produced in $e^+e^- \to q\bar{q}$ continuum events.

\begin{figure*}[ht!]
    \label{fig:F3}
    \begin{center}
    \includegraphics[height=0.29\textheight,width=0.495\linewidth]{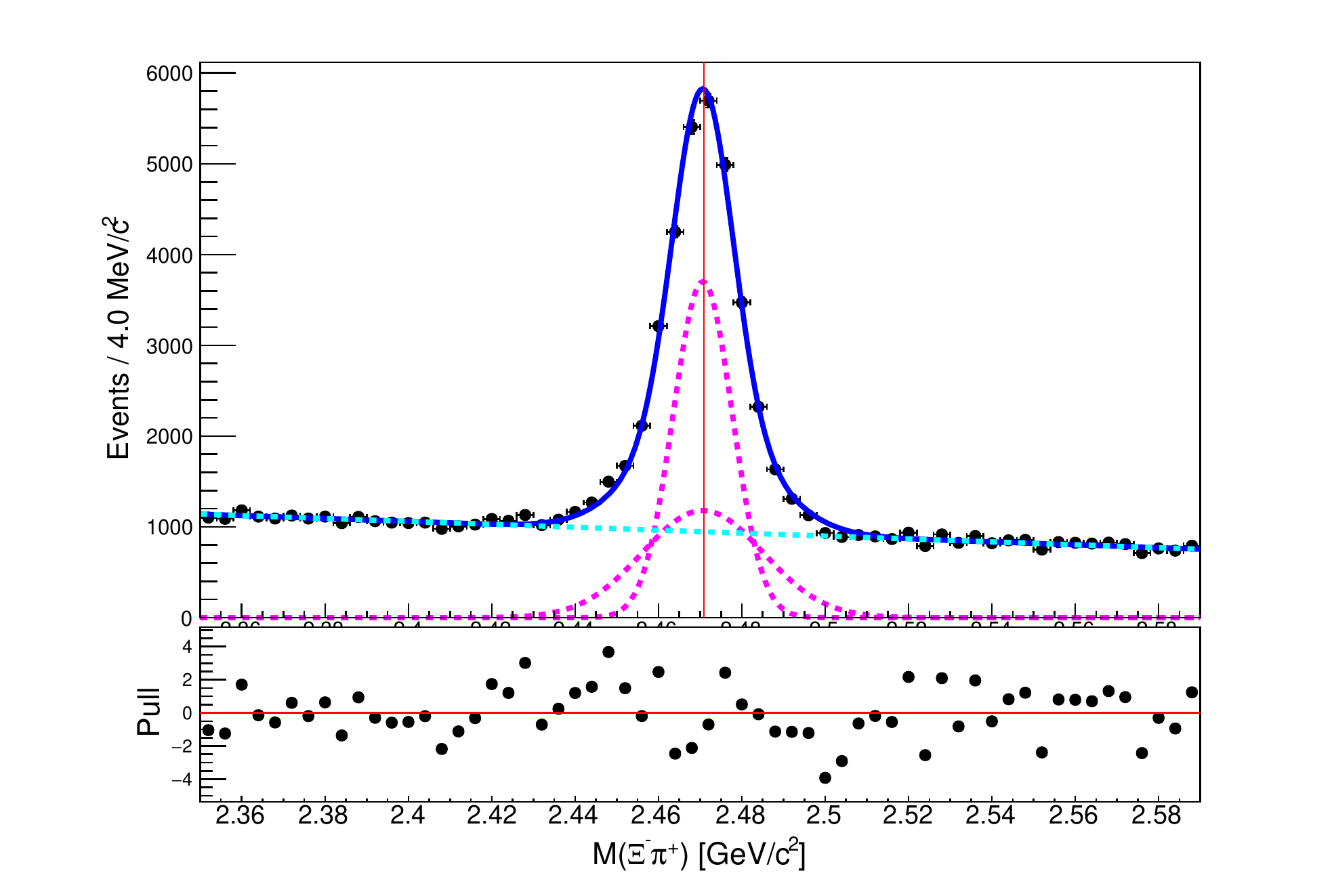}
    \includegraphics[height=0.29\textheight,width=0.495\linewidth]{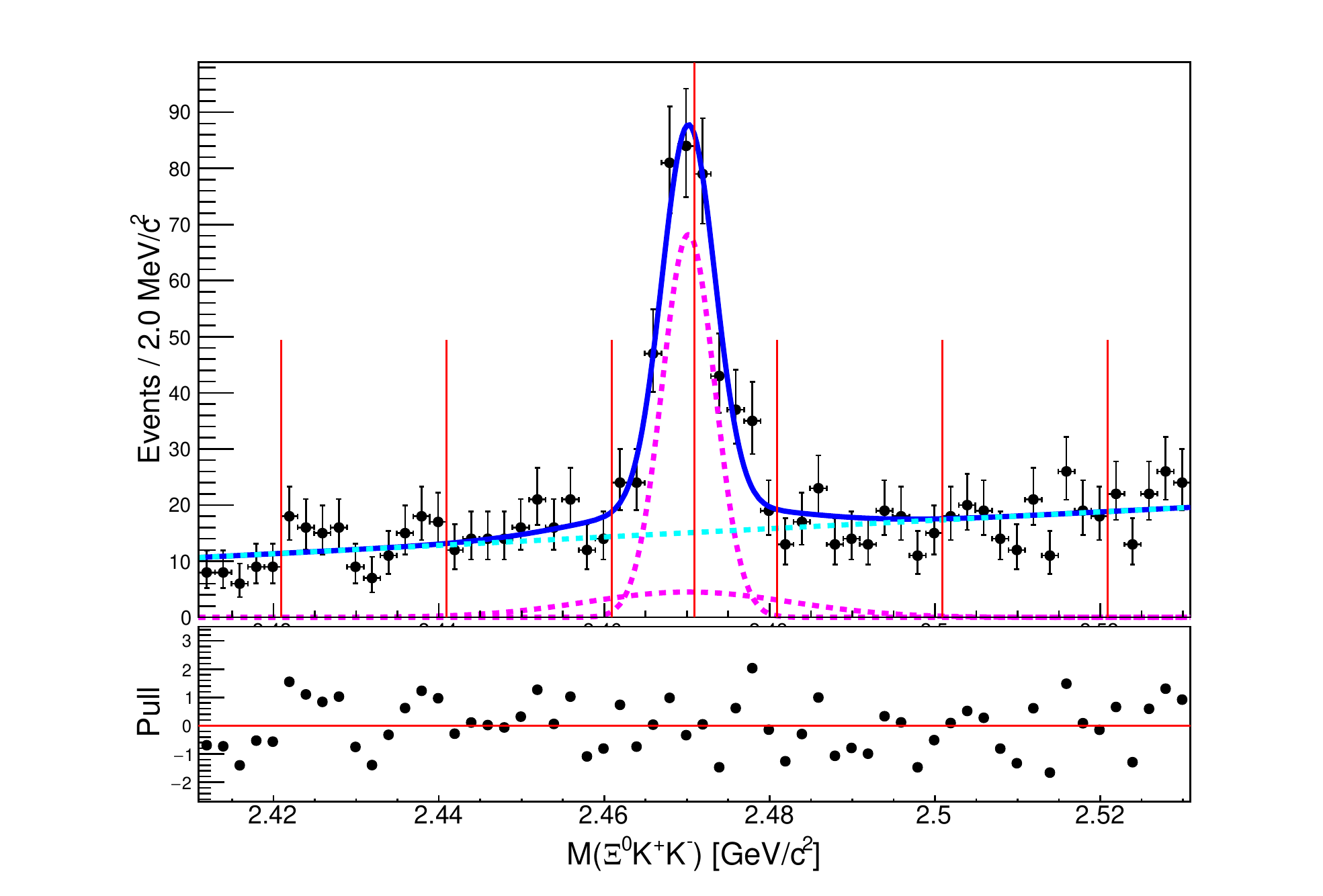}
    \end{center}
    \caption{Reconstructed invariant mass distributions for the normalization mode $\Xi_c^0 \to \Xi^- \pi^+$ (left) and the analysis mode $\Xi_c^0 \to \Xi^0 K^+ K^-$ (right) in the data sample (black) with an unbinned likelihood fit to a double Gaussian PDF (solid blue) using fixed MC resolutions with Chebyshev polynomial background (dashed cyan) and each individual Gaussian contribution to the signal PDF (dashed magenta). The pull distribution for each likelihood fit is included (bottom) adjacent to each reconstructed distribution. The selected signal, left, and right sidebands (marked in red half-lines) with respect to the nominal masses (marked in red full-lines) are described in the text.}
\end{figure*}

From the Monte-Carlo (MC), the PDF which best describes the $\Xi_c^0$ baryon is a double Gaussian with a primary, core, resolution and secondary resolution along with a second-order Chebyshev polynomial background. In \hyperref[fig:F3]{Fig. 3}, we plot the results of these unbinned maximum likelihood fits to the invariant mass distributions of the normalization channel, $\Xi_c^0 \to \Xi^- \pi^+$, and the signal channel, $\Xi_c^0 \to \Xi^0 K^+ K^-$, with fixed primary and secondary resolutions from MC simulations but with a free parameter yield into each Gaussian. We define the reconstruction efficiency into each mode as a ratio of signal yields between the reconstruction level and generator level for generic MC events containing a single generated $\Xi_c^0$. 

For the normalization mode, the primary and secondary resolutions are independently found to be $7.1\pm0.1$ MeV and $14.5\pm0.2$ MeV, respectively, which correspond to an RMS resolution of $9.7\pm0.1$ MeV and has a reconstruction efficiency $\epsilon_{\Xi^{-}\pi^{+}} = (7.04\pm0.05)\%$ with respect to the generated MC when resolutions are constrained. For the non-resonant and resonant signal modes, these resolutions are determined independently from $\Xi_c^0 \to \Xi^0 K^+ K^-$ MC generated with phase space distribution and $\Xi_c^0 \to \Xi^0 \phi (\to K^+ K^-)$ MC generated with a helicity amplitude distribution \cite{Lambdab,Lambdac,evtgen}, respectively. The difference in resolution between resonant and non-resonant decays was found to be negligible and are averaged with equal weighting. These average primary and secondary resolutions are found to be $3.19\pm0.02$ MeV and $12.37\pm0.21$ MeV, respectively, and correspond to an RMS resolution of $5.6\pm0.1$ MeV. Each signal mode MC sample, $\Xi_c^0 \to \Xi^0 K^+ K^-$ and $\Xi_c^0 \to \Xi^0 \phi (\to K^+ K^-)$, has statistically equivalent reconstruction efficiencies for their own resolution constraints given as $\epsilon_{\Xi^{0}K^{+}K^{-}} = (1.08\pm0.01)\%$ and $\epsilon_{\Xi^{0}\phi} = (1.09\pm0.01)\%$ respectively within a $\pm10$ MeV range of the $\Xi_c^0$ nominal mass \cite{PDG} which is approximately $3\sigma$ with respect to the primary resolution. 

From sideband samples in the $\Xi^0K^+K^-$ invariant mass distribution, we find that a significant number of $\phi \to K^+ K^-$ decays which are not from a resonant $\Xi_c^0$ are reconstructed in the $\pm 10$ MeV signal band. To account for these events, as well as the uniform combinatoral background, all reconstructed signal candidates in the signal band range of the $\Xi_c^0$ nominal mass \cite{PDG} are sideband-subtracted. This is done by subtraction of the scaled sideband candidates in the left and right sidebands of equal $\pm 10$ MeV width at central mass energies labelled in \hyperref[fig:F3]{Fig. 3} of $\pm 40$ MeV to the nominal mass with respect to the $\pm 10$ MeV signal band.

From the resulting likelihood fits in \hyperref[fig:F3]{Fig. 3} with resolutions constrained to the MC values previously defined, we find $n_{\Xi^+\pi^-} = 27186 \pm 475$ normalization mode candidates over the entire distribution range. Similarly, within the $\pm 10$ MeV range of the nominal mass, we find $n_{\rm cand.}\!=\!n_{\Xi^{0}\phi}\!+\!n_{\Xi^{0}K^{+}K^{-}}\!=\!311 \pm 23$ candidates into both signal modes by using a scaled sum of events in the sideband-subtracted distribution of the $\Xi^0K^+K^-$ invariant mass channel. Defining a signal statistical significance as $s/\delta(s)$ for signal yield $s$ and its uncertainty $\delta(s)$, our results correspond to a $13.5 \sigma$ statistical significance of these new Cabibbo-allowed $W$-exchange decays of $\Xi_c^0 \to \Xi^0 K^+ K^-$ including the resonant mode via $\phi \to K^+ K^-$.

\begin{figure}[h!]
    \label{fig:F4}
    \includegraphics[height=0.35\textheight,width=\linewidth]{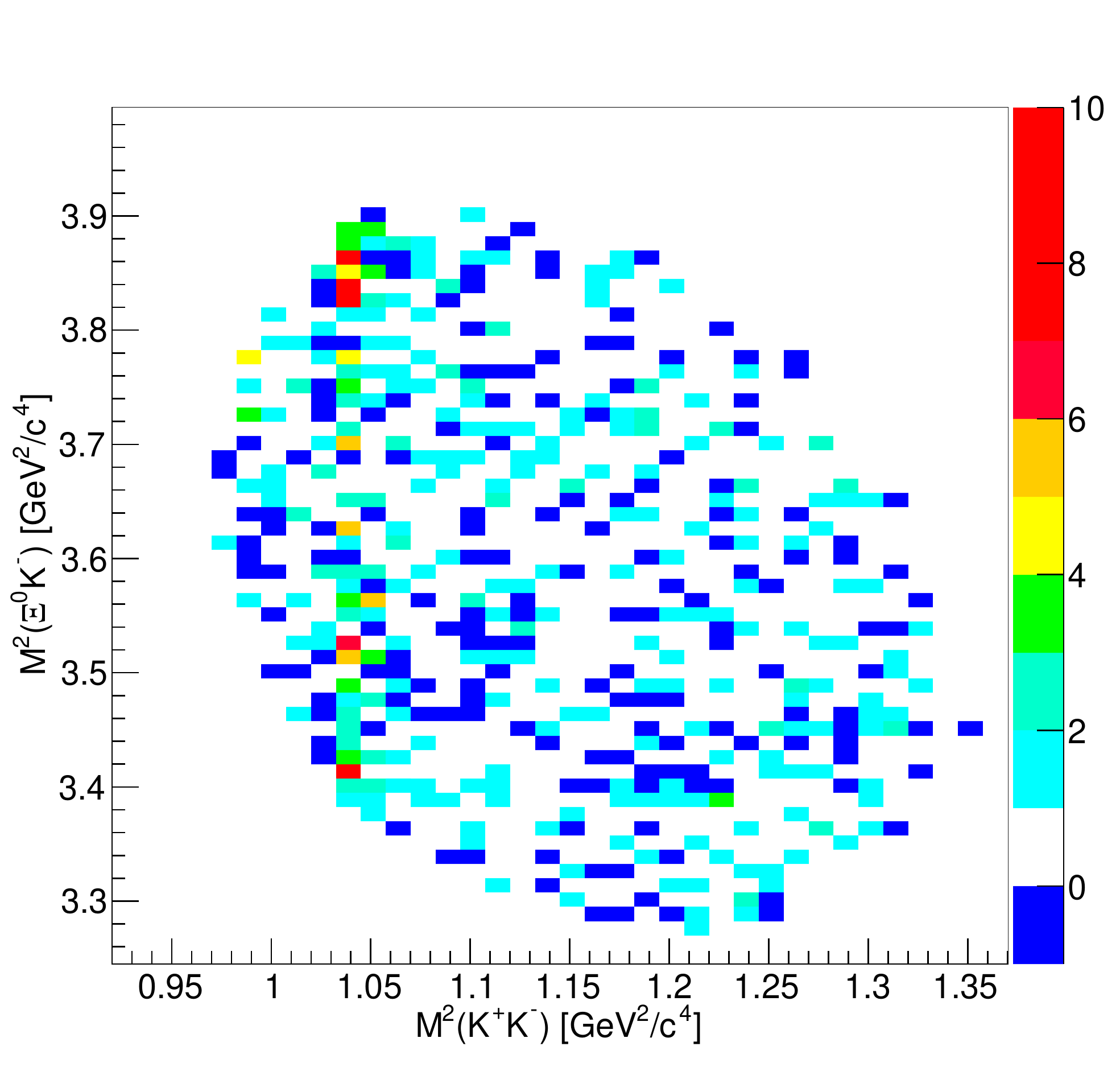}
    \caption{Dalitz plot distribution of the $\Xi_c^0 \to \Xi^0 K^+ K^-$ decays in the sideband-subtracted $\Xi_c^0$ signal region of the data sample, shown with square bins of $12.5$ MeV width.}
\end{figure}

The sideband-subtracted Dalitz plot is shown in \hyperref[fig:F4]{Fig. 4} with a mass-constrained $\Xi_c^0$ final state across the entire phase space. From this figure, we find a clear $\phi \to K^+K^-$ band but no evidence of other resonances. The $\phi$ resonance is found to be non-uniform due to the spin-polarization of the $\phi$ in the $1/2 \to 1/2 + 1$ resonant decay process $\Xi_c^0 \to \Xi^0 \phi$. This non-uniform substructure is specifically observed near $M^2(\Xi^0K^-)$ $\approx 3.85$ GeV$^2$ and $M^2(\Xi^0K^-)$ $\approx 3.425$ GeV$^2$ along the $\phi$ band at $M^2(K^+K^-)$ $\approx 1.04$ GeV$^2$. For this work, we define the branching ratios $\frac{\mathcal{B}(\rm mode)}{\mathcal{B}(\Xi_{c}^{0} \rightarrow \Xi^{-} \pi^{+})} = \frac{n_{\rm mode}}{\epsilon_{\rm mode}} / \frac{n_{\Xi^{-} \pi^{+}}}{\epsilon_{\Xi^{-} \pi^{+}}}$ with previously defined efficiencies. Using this definition, we study the fractions into these resonant and non-resonant modes using an amplitude analysis over this entire decay phase space. 

\section{\label{sec:S3}Amplitude Model}
The non-uniform contributions to the resonant substructure in the $\Xi_c^0 \to \Xi^0 \phi$ decay are best modelled using an amplitude analysis over the decay phase space. In this section, we provide basic forms used to model the spin-polarized, resonant amplitudes for $\phi \to K^+ K^-$ from the spin-$1/2$ $\Xi_c^0$ in an azimuthally symmetric plane. From the branching fractions into the resonant and non-resonant modes in \hyperref[fig:F4]{Fig. 4} and the efficiency corrected integration of amplitude intensities \cite{AmpTools}, we calculate the branching ratios $\frac{\mathcal{B}(\Xi_{c}^{0} \rightarrow \Xi^{0} \phi (\rightarrow K^{+}K^{-}))}{\mathcal{B}(\Xi_{c}^{0} \rightarrow \Xi^{-} \pi^{+})}$ and $\frac{\mathcal{B}(\Xi_{c}^{0} \rightarrow \Xi^{0} K^{+} K^{-})}{\mathcal{B}(\Xi_{c}^{0} \rightarrow \Xi^{-} \pi^{+})}$ as previously defined for the amplitude analysis.

\begin{figure}[ht!]
    \label{fig:F5}
    \includegraphics[height=0.225\textheight,width=0.95\linewidth]{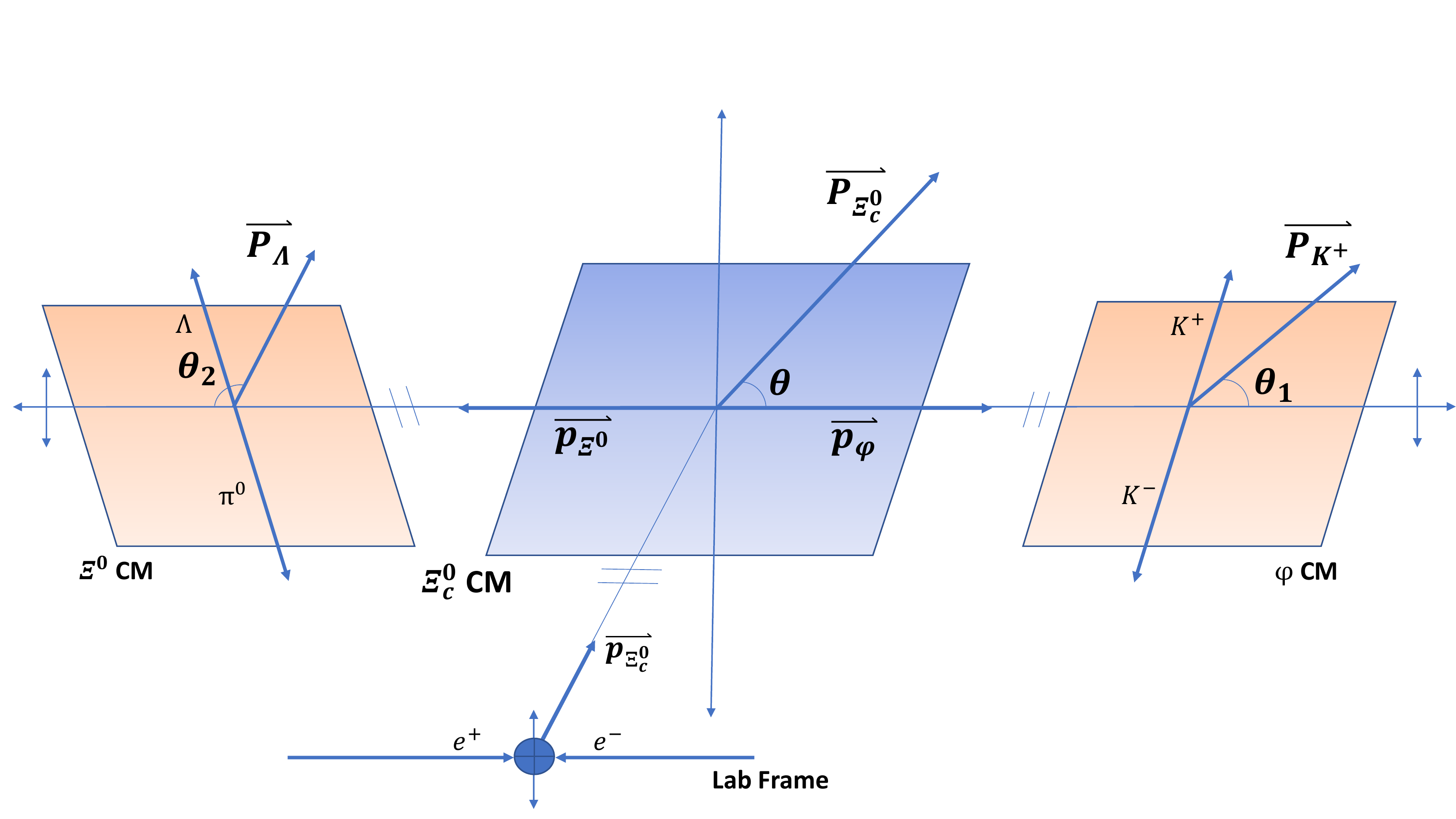}
    \caption{Azimuthally symmetric diagram of the spin-polarization angles in the resonant $\Xi_c^0 \to \Xi^0 \phi (\to K^+ K^-)$ decay.}
\end{figure}

It is known that a resonant decay amplitude is well modelled to first order using a Breit-Wigner amplitude \cite{suppressed,Physics}. To higher order, non-isotropic angular distributions contribute to the resonant amplitude due to the spin orientations of the final-state decay products. For the resonant $\Xi_c^0 \to \Xi^0 \phi (\to K^+ K^-)$ decay, the unit spin of the $\phi$ meson is polarized due to the $1/2$ spin of the heavy baryons. To describe this angular dependence of the polarized $\phi$, we study the polar polarization angles for the resonant $\Xi_c^0$ decay shown in \hyperref[fig:F5]{Fig. 5} with respect to the $\Xi_c^0$ momentum in the lab frame. In contrast, we assume the simple model that non-resonant $\Xi_c^0 \to \Xi^0K^+K^-$ are uniform, isotropic and are modeled using a constant amplitude over the entire phase space.
 
From \hyperref[fig:F5]{Fig. 5}, $\Xi_{c}^{0}$ decays resonantly into a $\phi$ with a polarization angle $\theta$ with respect to its polarization vector $\Vec{P_{\Xi_{c}^{0}}}$ in the lab frame. The polarization angle $\theta$ in this diagram is defined as the angle between the polarization vector and the resonant daughter momentum vector, $\Vec{p_{\phi}}$, in the $\Xi_{c}^{0}$ CM frame. Similarly, the resonant, spin-$1$ $\phi$ and spin-$1/2$ $\Xi^{0}$ decay with corresponding polarization angles $\theta_{1}$ and $\theta_{2}$ with respect to their momenta in the $\Xi_{c}^{0}$ CM frame and their daughter polarization vectors, $\Vec{P_{\Lambda}}$ and $\Vec{P_{K^{+}}}$, in their own respective CM frame.

\begin{figure*}[ht!]
    \label{fig:F6}
    \includegraphics[width=1.0\textwidth,height=0.25\textheight]{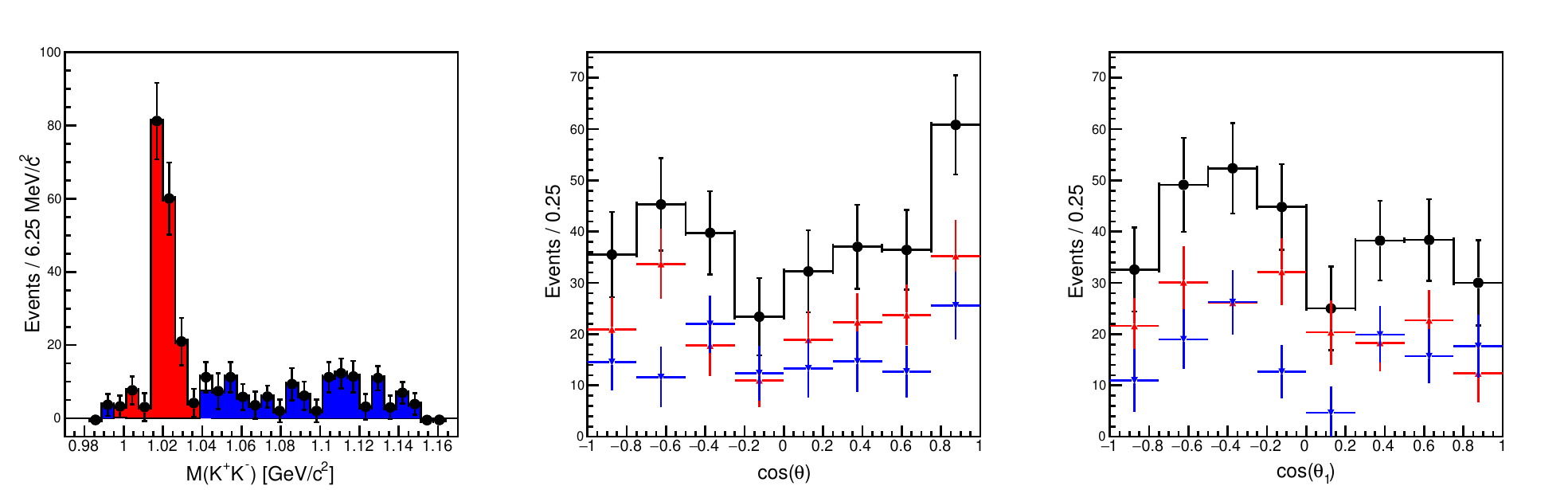} 
    \caption{Angular kinematics in the sideband-subtracted data sample for the $\cos(\theta)$ (middle) and $\cos(\theta_1)$ (right) distributions (black) in $\pm 20$ MeV range of the resonant nominal mass (red) and in non-resonant regions of phase space (blue) selected about the $K^{+} K^{-}$ invariant mass projection (left).}
\end{figure*}

After integration over $\theta_2$ in the Dalitz plot, the angular distribution of the resonant $\Xi_c^0$ decay depends effectively on the polarization angles of the $\Xi_c^0$ and the resonant $\phi$ coupled by their corresponding helicity amplitudes, $H_{\lambda_{\phi},\lambda_{\Xi^{0}}}$. We define this effective angular distribution of the resonant $\Xi_c^0$ decay amplitude as $A(\theta,\theta_1) \propto \mathbf{d^{1}}_{\lambda_{\phi},\lambda_{K^{+}}-\lambda_{K^{-}}}(\theta_{1}) H_{\lambda_{\phi},\lambda_{\Xi^{0}}} \mathbf{d^{1/2}}_{\lambda_{\Xi_{c}^{0}},\lambda_{\phi} - \lambda_{\Xi^{0}}}(\theta)$ for small-wigner D functions $\mathbf{d^{j}_{\lambda, \lambda_{1} - \lambda_{2}}}$ of a parent particle with spin $\mathbf{j}$ and helicity $\lambda$ decaying into two daughters with helicities $\mathbf{\lambda_{1}}$ \& $\mathbf{\lambda_{2}}$ \cite{Lambdac,Lambdab}. The kinematics in this effective angular distribution of our amplitude model are given in \hyperref[fig:F6]{Fig. 6} for $\cos(\theta)$ and $\cos(\theta_1)$ in $\pm 20$ MeV range of the resonant nominal mass and in non-resonant regions of phase space.

When fitting the Dalitz plot using this angular distribution, $A(\theta,\theta_1)$, due to the low statistics we reparameterize the sum of helicity contributions by absorbing constants into the overall normalization and eliminate any direct $H_{\lambda_{\phi},\lambda_{\Xi^{0}}}$ dependence which is outside the scope of this work. The resulting ratios of $H_{\lambda_{\phi},\lambda_{\Xi^{0}}}$ amplitudes are then free parameters in the amplitude model. Among these ratios, only the $\lambda_{\phi} = 0,$ $1$ terms contribute to the Dalitz plot which corresponds to the integrated distribution over both $\theta$ and $\theta_{2}$ and will only depend on $\theta_{1}$ directly.

From this, we conclude that the $\Xi_c^{0} \to \Xi^{0} K^{+} K^{-}$ amplitude from resonant, polarized $\phi \to K^{+}K^{-}$ decays can be described by combining this effective angular distribution, $A(\theta,\theta_1)$, with a Breit-Wigner amplitude, $V(E,M_{\phi},\Gamma)$. We then assert that this resonant amplitude is complex and coherent with the non-resonant constant amplitude, $\mathcal{A}_{KK}$,  across the Dalitz plot distribution described as $|\mathcal{A}_{KK} + \mathcal{A}_{\phi} \sum V(E,M_{\phi},\Gamma)A(\theta,\theta_1)|^{2}$ summed over all helicity states of $\Xi_c^0$, $\Xi^0$, and $\phi$ for normalization amplitudes $\mathcal{A}_{KK}$ and $\mathcal{A}_{\phi}$ of the non-resonant and resonant decays respectively. Alternate hypotheses for these assumptions are included in the systematic uncertainties.

\section{\label{sec:S4}Amplitude Analysis}
\begin{figure*}[ht!]
    \label{fig:F7}
    \includegraphics[width=1.0\textwidth,height=0.25\textheight]{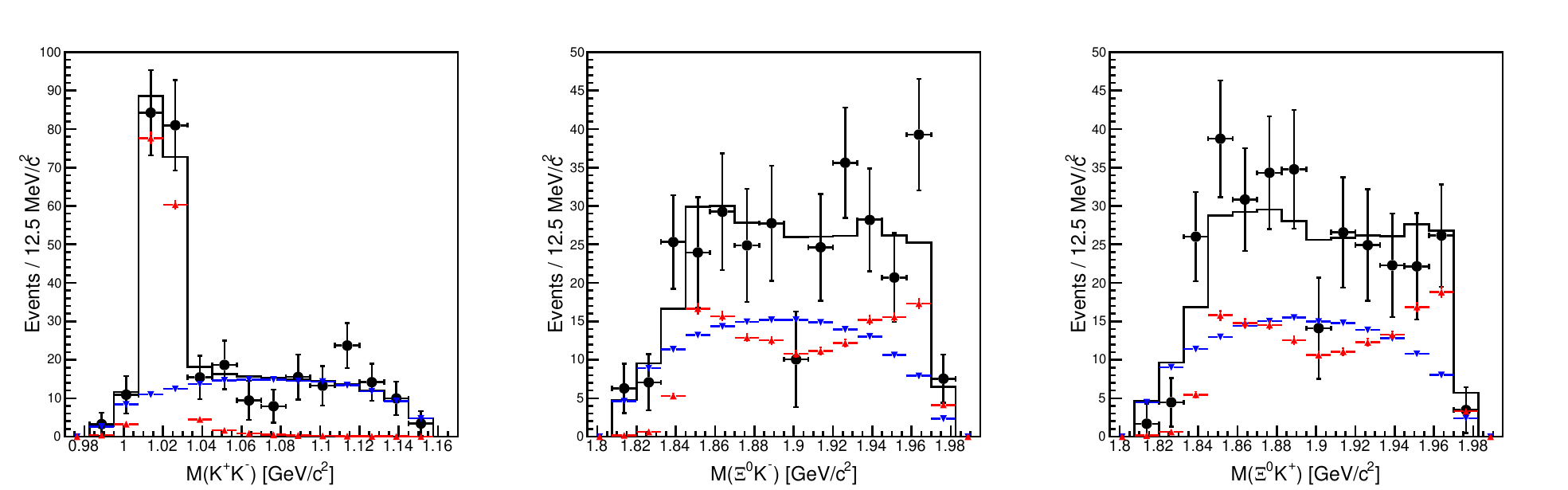}    
    \caption{Paired invariant mass projections of the amplitude intensities in the $\Xi_{c}^{0} \rightarrow \Xi^{0} K^{+} K^{-}$ decay via the resonant $\phi \rightarrow K^{+}K^{-}$ (red) and non-resonant (blue) modes in the sideband-subtracted data sample (black plot points) using the coherent sum of azimuthally symmetric amplitudes (solid black) as a model.}
\end{figure*}

We fit the Dalitz plot distribution in \hyperref[fig:F4]{Fig. 4} as a coherent sum of resonant and non-resonant amplitudes outlined in \hyperref[sec:S3]{Section III.} We then freely vary the helicity amplitude ratios of the resonant decay in the reparameterized angular distribution using only the natural width of the $\phi$ resonance, $\Gamma = 4.249$ MeV \cite{PDG}, as a constraint. The convolution of this natural width with the Gaussian resolution is included in the systematic uncertainties. The result of this unbinned maximum likelihood fit is shown in \hyperref[fig:F7]{Fig. 7} across each pair of invariant mass projections using the amplitude analysis software AmpTools (v.10.2) \cite{AmpTools}. From the normalized integration of each amplitude relative to the integral of their coherent sum, we measure branching fractions into the resonant and non-resonant modes as $(48.1 \pm 4.2) \%$ and $(51.9 \pm 4.2) \%$, respectively. In addition, we find that the measured mass of the $\phi$ meson, $M_{\phi} = 1019.62 \pm 0.25$ MeV, is in agreement with the current average value \cite{PDG}. 

These branching fractions for the resonant and non-resonant decay modes correspond to $n_{\Xi^{0}\phi} = 149 \pm 17$ and $n_{\Xi^{0}K^{+}K^{-}} = 161 \pm 18$ candidates, respectively. After varying all free parameters of our amplitude model \cite{AmpTools} and considering the reconstruction efficiencies across the Dalitz plot distribution from signal MC, we measure the precise resonant and non-resonant branching ratios into these new $W$-exchange $s\bar{s}$-popping decay modes:
\begin{eqnarray}
    \label{eq:E1}
    & \frac{\mathcal{B}(\Xi_{c}^{0} \rightarrow \Xi^{0} \phi (\rightarrow K^{+}K^{-}))}{\mathcal{B}(\Xi_{c}^{0} \rightarrow \Xi^{-} \pi^{+})} = 0.036 \pm 0.004\;(\rm stat.)\\
    \label{eq:E2}
    & \frac{\mathcal{B}(\Xi_{c}^{0} \rightarrow \Xi^{0} K^{+} K^{-})}{\mathcal{B}(\Xi_{c}^{0} \rightarrow \Xi^{-} \pi^{+})} = 0.039 \pm 0.004\;(\rm stat.)  
\end{eqnarray}

The branching ratios in \hyperref[eq:E1]{(1)} and \hyperref[eq:E2]{(2)} can easily be verified by one-dimensional analysis methods on the $\Xi^0 K^+ K^-$ invariant mass distribution in \hyperref[fig:F3]{Fig. 3}. For this validation, we plot the invariant mass of the $K^{+}K^{-}$ under the signal $\Xi_{c}^{0}$ band and weight the left and right sideband events appropriately, akin to the amplitude analysis. We then model the resonant $\phi \rightarrow K^{+}K^{-}$ in the $K^{+}K^{-}$ projection with a P-wave relativistic Breit-Wigner PDF convolved with a double Gaussian using constrained resolutions from signal MC studies. Using this analysis method we find statistically consistent results corresponding to branching fractions $(48.7 \pm 4.9) \%$ and $(51.3 \pm 4.9) \%$ into the resonant and non-resonant modes respectively. The corresponding minute differences between these measurements as branching ratios are consistent with a slight overestimation of the resonant mode observed during the MC analysis of this channel using one-dimensional methods. The methodological systematic uncertainty between the amplitude analysis and the one-dimensional analysis branching ratios is included in the total systematic uncertainty of the final branching ratio measurement. 

\section{\label{sec:S5}Systematic Uncertainties}
In \hyperref[tab:T1]{Table I}, we summarize all systematic uncertainties which impact our branching ratio measurements, summed in quadrature, to calculate the total systematic uncertainty. As the branching fraction is defined by the ratio of $\Xi_{c}^{0}\!\to\!(\Xi^{0}\!\to\!\Lambda^{0}\pi^{0})K^{+}K^{-}$ and $\Xi_{c}^{0}\!\to\!(\Xi^{-}\!\to\!\Lambda^{0}\pi^{-})\pi^{+}$, the detection efficiency relevant (DER) errors from particle identification and reconstruction common to both modes, such as $\Lambda$ finding and tracking, cancel. The remaining uncertainties are due to $\pi^0$ reconstruction ($\pm1.5 \%$) and PID ($\pm1.3 \%$), giving a total $\pm2.0\%$ when quadratically summed. These are estimated by comparing data and MC for dedicated calibration samples. In addition, we assign a systematic uncertainty of $^{+0.0}_{-1.6}\%$ and $^{+0.0}_{-3.3}\%$ to each respective branching ratio measurement due to the DER errors from multiple candidates in the signal distributions as a result of cloned hyperon reconstructions and tracks which originate far from the IP. This is calculated by comparing measurements from MC samples to the truth generated. 

In order to check our methodology, we compare the results found for the branching ratios with the amplitude analysis to those found using the simpler one-dimensional approach of fitting the Dalitz plot projections. We assign a systematic uncertainty of $^{+1.0}_{-0.0} \%$ and $^{+0.0}_{-0.9} \%$ to each respective branching ratio measurement due to variances between this amplitude analysis and the one-dimensional analysis measurements.

Our remaining systematic uncertainties are due to the MC statistics assumed for the resolutions as well as all model assumptions used throughout this analysis. We assign a collective systematic uncertainty of $^{+0.5}_{-0.1}\%$ to each branching ratio measurement due to the MC assumptions in our PDF models. This is calculated by varying and floating the values of the resolution constraints about the statistical uncertainty and summing these effects in quadrature. Similarly, we assign a systematic uncertainty of $^{+0.8}_{-0.0}\%$ to each branching ratio measurement due to the choice of PDF models used in our measurements of the $\Xi_c^0$ modes. This is calculated by finding the change in the result if we use a Gaussian PDF or a reduced fit range when modelling the reconstructed $\Xi_{c}^{0}$ invariant masses.

For the amplitude model assumptions, we assign a collective systematic uncertainty of $^{+3.8}_{-4.0}\%$ and $^{+3.3}_{-3.8}\%$ to each respective branching ratio measurement when quadratically summed. These are calculated by varying the model assumptions correlating to the natural width constraint \cite{PDG} and various measurements with alternative or ancillary amplitude models. These collective calculations include the following systematic effects: the quality of the phase space efficiencies and generated MC which is calculated by measuring these branching ratios with integrated efficiencies ($^{+0.0}_{-0.3}\%$ and $^{+0.3}_{-0.0}\%$); the effect of smearing the Breit-Wigner natural width by the reconstruction resolution ($^{+2.9}_{-4.0}\%$ and $^{+3.2}_{-2.5}\%$); the effect of using incoherent, non-interference models of the two amplitudes ($< \pm 0.1\%$); the effect of including contributions due to $a_0(980)$ and $f_0(980)$ mesons near threshold ($< \pm 0.1\%$); the assumption of azimuthal symmetry found by perturbing the model about small Euler angles ($\pm 0.1\%$); and the effect of direct helicity amplitude dependencies, the reparameterization, and the defined polarization angles calculated by systematically varying and eliminating these free parameters in our amplitude model ($^{+2.3}_{-0.5}\%$ and $^{+0.4}_{-2.8}\%$).

Summing all of these systematic uncertainties in quadrature, as shown in \hyperref[tab:T1]{Table I}, we assign a total systematic certainty of $^{+4.5}_{-4.8}\%$ and $^{+4.0}_{-5.5}\%$ for the two branching ratio measurements, respectively. 

\begin{table}[htp]
    \label{tab:T1}
    \begin{center}
    \renewcommand{\arraystretch}{1.3}
    \begin{tabular}{|c|c|c|}
    \hline \multicolumn{3}{|c|}{\textbf{Systematic Uncertainties}}\\
    \hline  & $\frac{\mathcal{B}(\Xi_{c}^{0} \rightarrow \Xi^{0} \phi (\rightarrow K^{+}K^{-}))}{\mathcal{B}(\Xi_{c}^{0} \rightarrow \Xi^{-} \pi^{+})}$ & $\frac{\mathcal{B}(\Xi_{c}^{0} \rightarrow \Xi^{0} K^{+} K^{-})}{\mathcal{B}(\Xi_{c}^{0} \rightarrow \Xi^{-} \pi^{+})}$\\
    \hline \multicolumn{3}{|c|}{Reconstruction}\\
    \hline $\pi^{0} \rightarrow \gamma \gamma$ \& PID & $\pm2.0$ & $\pm2.0$\\
    \hline \multicolumn{3}{|c|}{Multiple Candidates}\\
    \hline Clones \& Tracking & $^{+0.0}_{-1.6}$ & $^{+0.0}_{-3.3}$\\
    \hline \multicolumn{3}{|c|}{One-Dimensional Analysis}\\
    \hline $M(K^{+}K^{-})$ Yield & $^{+1.0}_{-0.0}$ & $^{+0.0}_{-0.9}$\\
    \hline \multicolumn{3}{|c|}{MC \& Model Assumptions}\\
    \hline MC Resolutions & $^{+0.5}_{-0.1}$ & $^{+0.5}_{-0.1}$ \\
    \hline PDF Models & $^{+0.8}_{-0.0}$ & $^{+0.8}_{-0.0}$\\
    \hline Amplitude Models & $^{+3.8}_{-4.0}$ & $^{+3.3}_{-3.8}$ \\
    \hline \multicolumn{3}{|c|}{\textbf{Total}}\\
    \hline & $^{+4.5}_{-4.8}$ & $^{+4.0}_{-5.5}$\\
    \hline
    \end{tabular}   
    \end{center}
    \caption{Contributions to the total systematic uncertainty of the branching ratio measurements, expressed as a percentage.}  
\end{table}

\section{\label{sec:S6}Conclusions}
Using the entire data sample of $980$ $\rm fb^{-1}$ integrated luminosity collected with the Belle detector \cite{PTEP}, we find a signal with statistical significance of $13.5\sigma$ for new $W$-exchange $s\bar{s}$-popping decay modes of $\Xi_{c}^{0}\!\to\!\Xi^0K^+K^-$ including resonant decays through $\phi\!\to\!K^+K^-$. Using an azimuthally symmetric amplitude model we find that among the $311 \pm 23$ candidates, $(48.1 \pm 4.2) \%$ decay resonantly through $\phi\!\to\!K^+ K^-$ while $(51.9 \pm 4.2) \%$ decay directly to $\Xi^0K^+K^-$. These yields are directly compared to the normalization mode $\Xi_c^0 \to \Xi^- \pi^+$ of yield $27186 \pm 475$ over the same data sample. From these measurements and the previously studied reconstruction efficiencies on signal MC, we report new branching ratios for these resonant and non-resonant $\Xi_c^0$ modes: 
\begin{eqnarray}
    & \frac{\mathcal{B}(\Xi_{c}^{0} \rightarrow \Xi^{0} \phi (\rightarrow K^{+}K^{-}))}{\mathcal{B}(\Xi_{c}^{0} \rightarrow \Xi^{-} \pi^{+})} \nonumber \\
    & = 0.036 \pm 0.004\;(\rm stat.) \pm 0.002\;(\rm syst.) \nonumber\\
    & \frac{\mathcal{B}(\Xi_{c}^{0} \rightarrow \Xi^{0} K^{+} K^{-})}{\mathcal{B}(\Xi_{c}^{0} \rightarrow \Xi^{-} \pi^{+})} \nonumber \\
    & = 0.039 \pm 0.004\;(\rm stat.) \pm 0.002\;(\rm syst.) \nonumber
\end{eqnarray}

The measurements of these $\Xi_c^0$ decay modes, which can only proceed via $W$-exchange together with $s\bar{s}$ production, add to our knowledge of the weak decay of charmed baryons. However, from the amplitude intensities in \hyperref[fig:F7]{Fig. 7}, we conclude it is unlikely that contributions from these resonant $\Xi_c^0 \to \Xi^0 \phi (\to K^+ K^-)$ decays will correlate to significant event excesses in the $\Xi^0 K^-$ reconstruction near $1.95$ GeV. Only minor cusping to the combinatorial background will be present in that region due to this decay. As a result of the slightly smaller branching fraction via the resonant $\Xi_c^0\to\Xi^0\phi$ decay, the apparent spin-polarization substructure is diluted due to the non-resonant fraction in this same region. This implies that future excited $\Omega$ searches may promisingly search this region of the $\Xi^0 K^-$ invariant mass after the inclusion of these new modes during MC studies. Despite the low statistics of these new modes at Belle, this study via an amplitude analysis provides necessary tools for studying more resonant amplitude features in multibody charmed baryon decays in the forthcoming high-luminosity era.

\medskip

\section{\label{sec:S7}Acknowledgements}

We thank the KEKB group for the excellent operation of the
accelerator; the KEK cryogenics group for the efficient
operation of the solenoid; and the KEK computer group, and the Pacific Northwest National
Laboratory (PNNL) Environmental Molecular Sciences Laboratory (EMSL)
computing group for strong computing support; and the National
Institute of Informatics, and Science Information NETwork 5 (SINET5) for
valuable network support.  We acknowledge support from
the Ministry of Education, Culture, Sports, Science, and
Technology (MEXT) of Japan, the Japan Society for the 
Promotion of Science (JSPS), and the Tau-Lepton Physics 
Research Center of Nagoya University; 
the Australian Research Council including grants
DP180102629, 
DP170102389, 
DP170102204, 
DP150103061, 
FT130100303; 
Austrian Science Fund (FWF);
the National Natural Science Foundation of China under Contracts
No.~11435013,  
No.~11475187,  
No.~11521505,  
No.~11575017,  
No.~11675166,  
No.~11705209;  
Key Research Program of Frontier Sciences, Chinese Academy of Sciences (CAS), Grant No.~QYZDJ-SSW-SLH011; 
the  CAS Center for Excellence in Particle Physics (CCEPP); 
the Shanghai Pujiang Program under Grant No.~18PJ1401000;  
the Ministry of Education, Youth and Sports of the Czech
Republic under Contract No.~LTT17020;
the Carl Zeiss Foundation, the Deutsche Forschungsgemeinschaft, the
Excellence Cluster Universe, and the VolkswagenStiftung;
the Department of Science and Technology of India; 
the Istituto Nazionale di Fisica Nucleare of Italy; 
National Research Foundation (NRF) of Korea Grant
Nos.~2016R1\-D1A1B\-01010135, 2016R1\-D1A1B\-02012900, 2018R1\-A2B\-3003643,
2018R1\-A6A1A\-06024970, 2018R1\-D1A1B\-07047294, 2019K1\-A3A7A\-09033840,
2019R1\-I1A3A\-01058933;
Radiation Science Research Institute, Foreign Large-size Research Facility Application Supporting project, the Global Science Experimental Data Hub Center of the Korea Institute of Science and Technology Information and KREONET/GLORIAD;
the Polish Ministry of Science and Higher Education and 
the National Science Center;
the Ministry of Science and Higher Education of the Russian Federation, Agreement 14.W03.31.0026; 
University of Tabuk research grants
S-1440-0321, S-0256-1438, and S-0280-1439 (Saudi Arabia);
the Slovenian Research Agency;
Ikerbasque, Basque Foundation for Science, Spain;
the Swiss National Science Foundation; 
the Ministry of Education and the Ministry of Science and Technology of Taiwan;
and the United States Department of Energy and the National Science Foundation.

\nocite{*}

\bibliography{McNeil_Xic0Xi0KK}

\providecommand{\noopsort}[1]{}\providecommand{\singleletter}[1]{#1}%
\begin{thebibliography}{14}%
\makeatletter
\providecommand \@ifxundefined [1]{%
 \@ifx{#1\undefined}
}%
\providecommand \@ifnum [1]{%
 \ifnum #1\expandafter \@firstoftwo
 \else \expandafter \@secondoftwo
 \fi
}%
\providecommand \@ifx [1]{%
 \ifx #1\expandafter \@firstoftwo
 \else \expandafter \@secondoftwo
 \fi
}%
\providecommand \natexlab [1]{#1}%
\providecommand \enquote  [1]{``#1''}%
\providecommand \bibnamefont  [1]{#1}%
\providecommand \bibfnamefont [1]{#1}%
\providecommand \citenamefont [1]{#1}%
\providecommand \href@noop [0]{\@secondoftwo}%
\providecommand \href [0]{\begingroup \@sanitize@url \@href}%
\providecommand \@href[1]{\@@startlink{#1}\@@href}%
\providecommand \@@href[1]{\endgroup#1\@@endlink}%
\providecommand \@sanitize@url [0]{\catcode `\\12\catcode `\$12\catcode
  `\&12\catcode `\#12\catcode `\^12\catcode `\_12\catcode `\%12\relax}%
\providecommand \@@startlink[1]{}%
\providecommand \@@endlink[0]{}%
\providecommand \url  [0]{\begingroup\@sanitize@url \@url }%
\providecommand \@url [1]{\endgroup\@href {#1}{\urlprefix }}%
\providecommand \urlprefix  [0]{URL }%
\providecommand \Eprint [0]{\href }%
\providecommand \doibase [0]{https://doi.org/}%
\providecommand \selectlanguage [0]{\@gobble}%
\providecommand \bibinfo  [0]{\@secondoftwo}%
\providecommand \bibfield  [0]{\@secondoftwo}%
\providecommand \translation [1]{[#1]}%
\providecommand \BibitemOpen [0]{}%
\providecommand \bibitemStop [0]{}%
\providecommand \bibitemNoStop [0]{.\EOS\space}%
\providecommand \EOS [0]{\spacefactor3000\relax}%
\providecommand \BibitemShut  [1]{\csname bibitem#1\endcsname}%
\let\auto@bib@innerbib\@empty
\bibitem [{\citenamefont {Chistov}\ \emph {et~al.}(2013)\citenamefont {Chistov}
  \emph {et~al.}}]{suppressed}%
  \BibitemOpen
  \bibfield  {author} {\bibinfo {author} {\bibfnamefont {R.}~\bibnamefont
  {Chistov}} \emph {et~al.} (\bibinfo {collaboration} {Belle Collaboration}),\
  }\bibfield  {title} {\bibinfo {title} {{First observation of
  Cabibbo-suppressed $\Xi_c^0$ decays}},\ }\href
  {https://doi.org/10.1103/PhysRevD.88.071103} {\bibfield  {journal} {\bibinfo
  {journal} {Phys. Rev. D}\ }\textbf {\bibinfo {volume} {88}},\ \bibinfo
  {pages} {071103} (\bibinfo {year} {2013})},\ \Eprint
  {https://arxiv.org/abs/1306.5947} {arXiv:1306.5947 [hep-ex]} \BibitemShut
  {NoStop}%
\bibitem [{\citenamefont {Zyla}\ \emph {et~al.}(2020)\citenamefont {Zyla} \emph
  {et~al.}}]{PDG}%
  \BibitemOpen
  \bibfield  {author} {\bibinfo {author} {\bibfnamefont {P.~A.}\ \bibnamefont
  {Zyla}} \emph {et~al.},\ }\bibfield  {title} {\bibinfo {title} {{(Particle
  Data Group)}},\ }\href {https://doi.org/10.1093/ptep/ptaa104} {\bibfield
  {journal} {\bibinfo  {journal} {Prog. Theor. Exp. Phys.}\ }\textbf {\bibinfo
  {volume} {2020}} (\bibinfo {year} {2020})},\ \bibinfo {note}
  {083C01}\BibitemShut {NoStop}%
\bibitem [{\citenamefont {Yelton}\ \emph {et~al.}(2018)\citenamefont {Yelton}
  \emph {et~al.}}]{excitedOm}%
  \BibitemOpen
  \bibfield  {author} {\bibinfo {author} {\bibfnamefont {J.}~\bibnamefont
  {Yelton}} \emph {et~al.} (\bibinfo {collaboration} {Belle Collaboration}),\
  }\bibfield  {title} {\bibinfo {title} {{Observation of an Excited $\Omega^-$
  Baryon}},\ }\href {https://doi.org/10.1103/PhysRevLett.121.052003} {\bibfield
   {journal} {\bibinfo  {journal} {Phys. Rev. Lett.}\ }\textbf {\bibinfo
  {volume} {121}},\ \bibinfo {pages} {052003} (\bibinfo {year} {2018})},\
  \Eprint {https://arxiv.org/abs/1805.09384} {arXiv:1805.09384 [hep-ex]}
  \BibitemShut {NoStop}%
\bibitem [{\citenamefont {Jia}\ \emph {et~al.}(2019)\citenamefont {Jia} \emph
  {et~al.}}]{Shen}%
  \BibitemOpen
  \bibfield  {author} {\bibinfo {author} {\bibfnamefont {S.}~\bibnamefont
  {Jia}} \emph {et~al.} (\bibinfo {collaboration} {Belle Collaboration}),\
  }\bibfield  {title} {\bibinfo {title} {{Search for $\Omega(2012)\to
  K\Xi(1530) \to K\pi\Xi$ at Belle}},\ }\href
  {https://doi.org/10.1103/PhysRevD.100.032006} {\bibfield  {journal} {\bibinfo
   {journal} {Phys. Rev. D}\ }\textbf {\bibinfo {volume} {100}},\ \bibinfo
  {pages} {032006} (\bibinfo {year} {2019})},\ \Eprint
  {https://arxiv.org/abs/1906.00194} {arXiv:1906.00194 [hep-ex]} \BibitemShut
  {NoStop}%
\bibitem [{\citenamefont {Gutsche}\ \emph {et~al.}(2013)\citenamefont {Gutsche}
  \emph {et~al.}}]{Lambdab}%
  \BibitemOpen
  \bibfield  {author} {\bibinfo {author} {\bibfnamefont {T.}~\bibnamefont
  {Gutsche}} \emph {et~al.},\ }\bibfield  {title} {\bibinfo {title}
  {{Polarization effects in the cascade decay $\Lambda_b \to \Lambda (\to
  p\pi^-) + J/\Psi (\to l^+l^-)$ in the covariant confined quark model}},\
  }\href {https://doi.org/10.1103/PhysRevD.88.114018} {\bibfield  {journal}
  {\bibinfo  {journal} {Phys. Rev. D}\ }\textbf {\bibinfo {volume} {88}},\
  \bibinfo {pages} {114018} (\bibinfo {year} {2013})},\ \Eprint
  {https://arxiv.org/abs/1309.7879} {arXiv:1309.7879 [hep-ph]} \BibitemShut
  {NoStop}%
\bibitem [{\citenamefont {Brodzicka}\ \emph {et~al.}(2012)\citenamefont
  {Brodzicka} \emph {et~al.}}]{PTEP}%
  \BibitemOpen
  \bibfield  {author} {\bibinfo {author} {\bibfnamefont {J.}~\bibnamefont
  {Brodzicka}} \emph {et~al.} (\bibinfo {collaboration} {Belle
  Collaboration}),\ }\bibfield  {title} {\bibinfo {title} {{Physics
  Achievements from the Belle Experiment}},\ }\href
  {https://doi.org/10.1093/ptep/pts072} {\bibfield  {journal} {\bibinfo
  {journal} {Prog. Theor. Exp. Phys.}\ }\textbf {\bibinfo {volume} {2012}},\
  \bibinfo {pages} {04D001} (\bibinfo {year} {2012})},\ \Eprint
  {https://arxiv.org/abs/1212.5342} {arXiv:1212.5342 [hep-ex]} \BibitemShut
  {NoStop}%
\bibitem [{\citenamefont {Abashian}\ \emph {et~al.}(2002)\citenamefont
  {Abashian} \emph {et~al.}}]{belle}%
  \BibitemOpen
  \bibfield  {author} {\bibinfo {author} {\bibfnamefont {A.}~\bibnamefont
  {Abashian}} \emph {et~al.} (\bibinfo {collaboration} {Belle Collaboration}),\
  }\bibfield  {title} {\bibinfo {title} {{The Belle Detector}},\ }\href
  {https://doi.org/10.1016/S0168-9002(01)02013-7} {\bibfield  {journal}
  {\bibinfo  {journal} {Nucl. Instrum. Meth. A}\ }\textbf {\bibinfo {volume}
  {479}},\ \bibinfo {pages} {117} (\bibinfo {year} {2002})}\BibitemShut
  {NoStop}%
\bibitem [{\citenamefont {Bevan}\ \emph {et~al.}(2014)\citenamefont {Bevan}
  \emph {et~al.}}]{Physics}%
  \BibitemOpen
  \bibfield  {author} {\bibinfo {author} {\bibfnamefont {A.}~\bibnamefont
  {Bevan}} \emph {et~al.} (\bibinfo {collaboration} {BaBar, Belle
  Collaborations}),\ }\bibfield  {title} {\bibinfo {title} {{The Physics of the
  B Factories}},\ }\href {https://doi.org/10.1140/epjc/s10052-014-3026-9}
  {\bibfield  {journal} {\bibinfo  {journal} {Eur. Phys. J. C}\ }\textbf
  {\bibinfo {volume} {74}},\ \bibinfo {pages} {3026} (\bibinfo {year}
  {2014})},\ \Eprint {https://arxiv.org/abs/1406.6311} {arXiv:1406.6311
  [hep-ex]} \BibitemShut {NoStop}%
\bibitem [{\citenamefont {Kuhr}\ \emph {et~al.}(2019)\citenamefont {Kuhr} \emph
  {et~al.}}]{basf2}%
  \BibitemOpen
  \bibfield  {author} {\bibinfo {author} {\bibfnamefont {T.}~\bibnamefont
  {Kuhr}} \emph {et~al.} (\bibinfo {collaboration} {Belle II Framework Software
  Group}),\ }\bibfield  {title} {\bibinfo {title} {{The Belle II Core
  Software}},\ }\href {https://doi.org/10.1007/s41781-018-0017-9} {\bibfield
  {journal} {\bibinfo  {journal} {Comput. Softw. Big Sci.}\ }\textbf {\bibinfo
  {volume} {3}},\ \bibinfo {pages} {1} (\bibinfo {year} {2019})},\ \Eprint
  {https://arxiv.org/abs/1809.04299} {arXiv:1809.04299 [physics.comp-ph]}
  \BibitemShut {NoStop}%
\bibitem [{\citenamefont {Gelb}\ \emph {et~al.}(2018)\citenamefont {Gelb} \emph
  {et~al.}}]{b2bii}%
  \BibitemOpen
  \bibfield  {author} {\bibinfo {author} {\bibfnamefont {M.}~\bibnamefont
  {Gelb}} \emph {et~al.} (\bibinfo {collaboration} {Belle II Collaboration}),\
  }\bibfield  {title} {\bibinfo {title} {{B2BII: Data Conversion from Belle to
  Belle II}},\ }\href {https://doi.org/10.1007/s41781-018-0016-x} {\bibfield
  {journal} {\bibinfo  {journal} {Comput. Softw. Big Sci.}\ }\textbf {\bibinfo
  {volume} {2}},\ \bibinfo {pages} {9} (\bibinfo {year} {2018})},\ \Eprint
  {https://arxiv.org/abs/1810.00019} {arXiv:1810.00019 [hep-ex]} \BibitemShut
  {NoStop}%
\bibitem [{\citenamefont {Krohn}\ \emph {et~al.}(2020)\citenamefont {Krohn}
  \emph {et~al.}}]{treefitter}%
  \BibitemOpen
  \bibfield  {author} {\bibinfo {author} {\bibfnamefont {J.-F.}\ \bibnamefont
  {Krohn}} \emph {et~al.},\ }\bibfield  {title} {\bibinfo {title} {{Global
  decay chain vertex fitting at Belle II}},\ }\href
  {https://doi.org/https://doi.org/10.1016/j.nima.2020.164269} {\bibfield
  {journal} {\bibinfo  {journal} {Nucl. Instrum. Meth. A}\ }\textbf {\bibinfo
  {volume} {976}},\ \bibinfo {pages} {164269} (\bibinfo {year}
  {2020})}\BibitemShut {NoStop}%
\bibitem [{\citenamefont {Gutsche}\ \emph {et~al.}(2015)\citenamefont {Gutsche}
  \emph {et~al.}}]{Lambdac}%
  \BibitemOpen
  \bibfield  {author} {\bibinfo {author} {\bibfnamefont {T.}~\bibnamefont
  {Gutsche}} \emph {et~al.},\ }\bibfield  {title} {\bibinfo {title}
  {{Semileptonic decay $\Lambda_b \to \Lambda_c + \tau^- + \bar{\nu_\tau}$ in
  the covariant confined quark model}},\ }\href
  {https://doi.org/10.1103/PhysRevD.91.074001} {\bibfield  {journal} {\bibinfo
  {journal} {Phys. Rev. D}\ }\textbf {\bibinfo {volume} {91}},\ \bibinfo
  {pages} {074001} (\bibinfo {year} {2015})},\ \bibinfo {note} {[Erratum:
  Phys.Rev.D 91, 119907 (2015)]},\ \Eprint {https://arxiv.org/abs/1502.04864}
  {arXiv:1502.04864 [hep-ph]} \BibitemShut {NoStop}%
\bibitem [{\citenamefont {Lange}(2001)}]{evtgen}%
  \BibitemOpen
  \bibfield  {author} {\bibinfo {author} {\bibfnamefont {D.~J.}\ \bibnamefont
  {Lange}},\ }\bibfield  {title} {\bibinfo {title} {{The EvtGen Particle Decay
  Simulation Package}},\ }\href {https://doi.org/10.1016/S0168-9002(01)00089-4}
  {\bibfield  {journal} {\bibinfo  {journal} {Nucl. Instrum. Meth. A}\ }\textbf
  {\bibinfo {volume} {462}},\ \bibinfo {pages} {152} (\bibinfo {year}
  {2001})}\BibitemShut {NoStop}%
\bibitem [{\citenamefont {Matevosyan}\ \emph {et~al.}(2019)\citenamefont
  {Matevosyan}, \citenamefont {Mitchell},\ and\ \citenamefont
  {Shepherd}}]{AmpTools}%
  \BibitemOpen
  \bibfield  {author} {\bibinfo {author} {\bibfnamefont {H.}~\bibnamefont
  {Matevosyan}}, \bibinfo {author} {\bibfnamefont {R.}~\bibnamefont
  {Mitchell}},\ and\ \bibinfo {author} {\bibfnamefont {M.}~\bibnamefont
  {Shepherd}},\ }\href@noop {} {\bibinfo {title} {Amptools}},\ \bibinfo
  {howpublished} {\url{https://github.com/mashephe/AmpTools/wiki}} (\bibinfo
  {year} {2016-2019})\BibitemShut {NoStop}%
\end{thebibliography}%

\end{document}